\documentclass[aps,pra,twocolumn,letterpaper,groupaddress,10pt]{revtex4-2}
\usepackage{amssymb,amsthm,amsmath,amsfonts}
\usepackage{graphicx,ulem,enumerate,bbm,mathptmx}
\usepackage[pdftex,dvipsnames,usenames]{xcolor}
\usepackage[colorlinks=true,urlcolor=blue,citecolor=blue,linkcolor=blue]{hyperref}

\begin{document}

\title{Entanglement-assisted quantum speedup: Beating local quantum speed limits}

\author{Farha Yasmin}
	\affiliation{Theoretical Quantum Science, Institute for Photonic Quantum Systems (PhoQS), Paderborn University, Warburger Stra\ss{}e 100, 33098 Paderborn, Germany}

\author{Jan Sperling}
	\email{jan.sperling@upb.de}
	\affiliation{Theoretical Quantum Science, Institute for Photonic Quantum Systems (PhoQS), Paderborn University, Warburger Stra\ss{}e 100, 33098 Paderborn, Germany}

\date{\today}

\begin{abstract}
	Research in quantum information science aims to surpass the scaling limitations of classical information processing.
	From a physicist's perspective, performance improvement involves a physical speedup in the quantum domain, achieved by dynamically exploiting quantum correlations.
	In this study, speed limits in interacting quantum systems are derived by comparing the rates of change in actual quantum dynamics with the quasi-classical evolution confined to the manifold of non-entangled separable states.
	The utility of the resulting bounds on entanglement-assisted speedup is demonstrated on bipartite qubit systems, bipartite qudit systems, as well as a complex multimode systems.
	Specifically, the proposed speed limits provide a tight bound on quantum speed advantage, including a quantum gain that can scale exponentially with the system's size.
	Extensions of the results to open systems and measurable witnesses are discussed.
\end{abstract}

\maketitle


\section{Introduction}

	To a large extent, the rapid growth of quantum information science originates from the promise to carry out computational tasks much faster than any classical computer.
	This is accomplished by harnessing the structure of quantum physics and exploiting quantum effects, such as entanglement, as a resource \cite{P12,CG19,A22}.
	Computational-complexity-theoretic advantages can, for example, be found in Shor's algorithm \cite{S94}, utilizing the power of the quantum Fourier transform, and Grover's algorithm \cite{G96}.
	However, characterizing quantum advantages presents a major challenge, such as assessing entanglement, which constitutes an NP-hard problem \cite{G03,I07}.
	In addition, found speedups are commonly expressed from a computer scientist's perspective, and providing a physical picture of this superior operation often constitutes a hurdle.

	Undoubtedly, entanglement is vital for many tasks in quantum information theory \cite{HHHH09}.
	Yet, the origin of this quantum correlation lies in fundamental considerations about the unique features of quantum physics \cite{EPR35,B35,B64}.
	However, merging both computational and physical aspects by quantifying the entanglement-assisted speedup of the dynamics in quantum correlated systems constitutes an open issue to date.

	The general study of the speed of quantum processes in physical terms is a vivid field of research \cite{DC17}.
	To this end, universal quantum speed limits have been established, also relating to energy-time uncertainty relations and the minimal time required for a quantum process \cite{MT45,AB61}.
	In general, bounding the rate of change turned out to be a useful tool to characterize the quantum evolution in open and closed systems;
	see, e.g., Refs. \cite{ML98,AA04,LT09,Y10,TEDM13,DL13,CEPH13,LXZ15,MTW16,WY18,CPBM18,FSS19,TP22,K22}.
	This further extends to recent studies of the operational quantum information flow via correlation functions \cite{SLZYL20,CHC22}.

	From Schr\"odinger's equation and related equations of motions, the quantum speed may be derived, and the resulting bound is determined by the Hamiltonian's spectrum \cite{DC17}.
	First attempts to relate such energetic considerations to quantum-computational aspects of speedups have been put forward, too \cite{F05,J17}.
	For example, the number of quantum gate operations per unit of time can be estimated by such a relation \cite{ML98,L00}.
	Still, a general relation for how the physical propagation in time relates to computational quantum advantages remains an unsolved problem, including a lack of a generally applicable framework to address such a question in the first place.

	Several applications and properties were found to demonstrate the practical usefulness of studying quantum speed limits.
	For example, quantum speed limits have been recently considered when restricted to linear subspaces \cite{AM22}, modeling inaccessible resources.
	Also, the order of an interaction can be related to the resulting speed of the buildup of entanglement \cite{GLM03}.
	In general, the overlap of an evolving state to another state may be assessed to quantify the approach towards an entangled target state \cite{T22}.
	Other geometric aspects of quantum states and their relation to quantum speed limits have been explored in Refs. \cite{AA90,V92,JK10,Z12}.
	Furthermore, the transition $\hbar\to0$ led to interesting insights into classical versus quantum speed limits \cite{BGK21}.
	The general quantum nature of quantum speed limits and its connection with classical speed limits have been discussed \cite{OO18,GNGCG22}.
	In relation to bath-induced classical-quantum transitions, experiments have, for example, demonstrated that systems interacting with a non-Markovian environment may evolve faster than those coupled to a memory-less bath \cite{CYPCOD15}.
	Further applications of quantum speed limits can be found in quantum metrology \cite{F12,GLM11,TEDM13,ZPK11,BC17}.
	The optimal control of quantum systems also benefits from quantifying the maximal rate of change of quantum states \cite{CMCFMGS09}, including related experiments \cite{BVMHACFGMM12}.
	Similar concepts apply to optimized operations in quantum thermodynamics, such as quantum battery charging tasks \cite{BVMG15,CPBCGVM17}.

	Somewhat separated from the research on general speed limits, the evolution of entanglement has been studied;
	see, e.g., Refs. \cite{WBPS06,FABWR07,TMB08,I09,G10,LZWLHTY17}.
	For instance, this led to notions like sudden death (and sudden birth) of entanglement \cite{YE04,YE09}.
	In general, input-output maps can be characterized according to their entangling capabilities to quantify the generation of output entanglement \cite{ZHHH01,GG12,MR13,SV15}.
	Beyond that, specific equations of motion render it possible to dynamically probe entanglement \cite{KMTKAB08,SW17,SW20}.
	Also, found relations between quantum causality and entanglement are interesting from a fundamental perspective \cite{ABCFGB15,B16}.
	However, entanglement is mostly discussed at each point in time separately---not actually addressing the entangling power of the dynamics itself---or is studied in terms of input-output relations between the initial and final time---missing out on the entanglement for all intermediate times.
	Importantly, a relation of entanglement dynamics to quantum speed limits has not been established thus far.

	In this contribution, we formulate quantum speed limits for arbitrary processes when restricted to non-entangled---i.e., separable \cite{W89}---states.
	This is achieved by applying nonlinear variational principles for constraining any quantum process to the manifold of  separable states (Sec. \ref{sec:Derivation}).
	The comparison of the thereby obtained local quantum speed limit with the actual rate of change of a given process renders it possible to quantify the entanglement-assisted speedup.
	Examples that relate to spin-correlated systems to implement fundamental quantum gates, evolutions near a ground state for ultracold systems, and multimode nonlinear optical processes are characterized in this fashion (Sec. \ref{sec:application}).
	This includes providing the scaling of the entanglement-caused speedup as a function of the dimension and number of parties of the interacting quantum system.
	Furthermore, measurable criteria in terms of two-time correlation functions are derived together with generalizations to open quantum dynamics (Sec. \ref{sec:PropertiesGeneralizations}).
	Thus, a universal and powerful methodology is formulated to assess the physical speedup that is caused by entanglement in interacting quantum systems.


\section{Derivation}
\label{sec:Derivation}

	In this section, we derive our methodology.
	Firstly, concepts for obtaining quantum speed limits, regardless of the entanglement properties, are discussed in Sec. \ref{subsec:QSLsingle}.
	Recently devised methods to characterize entanglement of states and the evolution are then introduced in Sec. \ref{subsec:QSLmultiple}.
	Eventually, both approaches are harnessed to determine separable quantum speed limits in systems with an arbitrary number of parties.

\subsection{Quantum speed limits for a single system}
\label{subsec:QSLsingle}

	Here, we begin with general quantum speed limits.
	While derivations similar to the one shown here can be found in previous works (see, e.g., Ref. \cite{DC17} for an overview), the specific calculation is useful as it serves as template for quantum speed limits when restricted to the manifold of separable states.

	A generic density operator for a mixed state, written as $\hat\rho=\sum_{n}p_n|\psi_n\rangle\langle\psi_n|$, evolves in a closed system according to the von~Neumann equation,
	\begin{equation}
		\label{eq:vonNeumann}
		\partial_t \hat\rho
		=\frac{1}{i\hbar}[\hat H,\hat\rho],
	\end{equation}
	with $\hat H$ being the Hamiltonian that governs the quantum dynamics.
	The total rate of change is given by $\|\partial_t\hat\rho\|_1$, where we chose to employ the trace norm, also known as one norm.
	(Recall that the trace norm for self-adjoined operators is given by the sum of the modulus of eigenvalues of said operator.)

	The maximally possible rate of change---the speed limit as used here---is then determined by the supremum of the expression $\|\partial_t\hat\rho\|_1$.
	Because of convexity and the fact that $\partial_t p_n=0$ holds true for processes that obey Eq. \eqref{eq:vonNeumann}, we may restrict ourselves to pure states.
	Now, the quantum speed limit reads
	\begin{equation}
		\label{eq:DefQSL}
		\mathsf{QSL}
		=\sup_{|\psi\rangle}\left\|
			\partial_t(|\psi\rangle\langle\psi|)
		\right\|_1
		=\sup_{|\psi\rangle}\left\|
			\frac{1}{i\hbar}\big[\hat H,|\psi\rangle\langle\psi|\big]
		\right\|_1.
	\end{equation}

	For determining the trace norm, one can firstly determine the eigenvalues $\gamma$ of the commutator, which is given by the Hermitian operator $\hat C=|\eta\rangle\langle\psi|+|\psi\rangle\langle\eta|$, where $|\eta\rangle=(i\hbar)^{-1}\hat H|\psi\rangle$.
	Thereby, one straightforwardly obtains the sought-after eigenvalues,
	\begin{equation}
		\label{eq:MaxEigenvalueCommutator}
		\gamma=\pm\frac{1}{\hbar}\sqrt{
			\langle\psi|(\Delta\hat H)^2|\psi\rangle
		},
	\end{equation}
	with $\Delta\hat H=\hat H-E\hat 1$ and $E=\langle\psi|\hat H|\psi\rangle$.
	Please note that it can be easily seen that all other eigenvalues are $0$---thus, not contributing to the norm.
	The non-zero eigenvalues $\gamma$ of the commutator  include the square root of the energy variance and have to be maximized to determine the quantum speed limit,
	\begin{equation}
		\label{eq:QSLviacommutatoreigenvalues}
		\mathsf{QSL}=2\sup_{|\psi\rangle}|\gamma|,
	\end{equation}
	according to Eq. \eqref{eq:DefQSL}.

	Further, the maximal expectation value of $(\Delta\hat H)^2$ is the maximal eigenvalue of this operator, as obtained by the eigenvalue equation
	\begin{equation}
		(\Delta\hat H)^2|\psi\rangle=(\hbar\gamma)^2|\psi\rangle.
	\end{equation}
	For solving this equation, suppose the spectral decomposition $\hat H=\sum_n E_n|n\rangle\langle n|$, implying $(\Delta\hat H)^2=\sum_n(E_n-E)^2|n\rangle\langle n|$.
	In the non-degenerate case, we find that the eigenvectors $|\psi\rangle=|n\rangle$ result in the variances $\gamma^2=0$, with $E=\langle n|\hat H|n\rangle=E_n$.
	On the other hand, we have degenerate eigenvalues if $(E_n-E)^2=(E_m-E)^2$ holds true for $m\neq n$, meaning that $E_n=E_m$ or $E=(E_m+E_n)/2$ applies.
	In addition to the previous eigenvalues, yielding $\gamma^2=0$, we therefore get the degenerate eigenvalues
	\begin{equation}
		(\hbar\gamma)^2=\left(\frac{E_m-E_n}{2}\right)^2.
	\end{equation}
	This means that the maximal variance is $\langle (\Delta\hat H)^2\rangle=([E_{\max}-E_{\min}]/2)^2$, where $E_{\max}$ and $E_{\min}$ respectively are the maximal and minimal eigenvalue of $\hat H$.

	Consequently, we have $\hbar\gamma=\pm(E_{\max}-E_{\min})/2$ as the maximal and minimal eigenvalue of the commutator.
	From that and Eq. \eqref{eq:QSLviacommutatoreigenvalues}, we finally find
	\begin{equation}
		\label{eq:QSL}
		\mathsf{QSL}
		=\frac{E_{\max}-E_{\min}}{\hbar}
	\end{equation}
	as the sought-after quantum speed limit of the process described by the Hamiltonian $\hat H$.

	Please note that the minimal and maximal values exist, $E_{\min}=\inf_n E_n>-\infty$ and $E_{\sup}=\sup_n E_n<\infty$, when the Hamiltonian is a bounded operator.
	Furthermore, in finite-dimensional (thus, compact) spaces, there exist eigenstates such that $\hat H|\min\rangle=E_{\min}|\min\rangle$ and $\hat H|\max\rangle=E_{\max}|\max\rangle$ hold true.
	A balanced superposition of those states yields the maximal energy variance, $([E_{\max}-E_{\min}]/2)^2$, showing that the bound $\mathsf{QSL}$ is tight in that scenario.

	Since the evolution of the system is controlled by the Hamiltonian's eigenvalues, it is sensible that the maximally possible rate of change is directly connected to the largest energy difference possible.
	This simple formula \eqref{eq:QSL} for the quantum speed limit shall be compared to the analogous speed limit when restricting to non-entangled states to assess quantum speedups caused by entangling interactions.

\subsection{Bounding separable speeds in multipartite systems}
\label{subsec:QSLmultiple}

\subsubsection{Preliminaries}

	We now put forward the idea to obtain speed limits for dynamics restricted to the manifold of separable states.
	This constitutes the main result of this work and serves as the basis for demonstrating physical exponential speedup, in addition to other results discussed in the remainder of this work.

	A separable---i.e., classically correlated---quantum state in an $N$-partite quantum state can be expressed in the form \cite{W89} 
	\begin{equation}
		\hat\rho_\mathrm{sep}=\sum_{n}p_n|a_{1,n},\ldots,a_{N,n}\rangle\langle a_{1,n},\ldots,a_{N,n}|,
	\end{equation}
	where $|a_{1,n},\ldots,a_{N,n}\rangle=\bigotimes_{j=1}^N|a_{j,n}\rangle$ denote product states that are mixed according to the probabilities $p_n$.

	For arbitrary Hamiltonians, the evolution confined to the set of separable states was derived in Refs. \cite{SW17,SW20}.
	Like in the previous single-party case, it was shown that $\partial_t p_n=0$ holds true in closed systems \cite{SW20}, allowing one to restrict to pure product states, $|a_1,\ldots,a_N\rangle$.

	Akin to Eq. \eqref{eq:vonNeumann}, the subsystem dynamics of a separable ensemble follows a von~Neumann-type equation \cite{SW17},
	\begin{equation}
		\label{eq:SepvNeumann}
		\partial_t\left(|a_j\rangle\langle a_j|\right)
		=\frac{1}{i\hbar}[\hat H_{a_1,\ldots,a_{j-1},a_{j+1},\ldots,a_N},|a_j\rangle\langle a_j|],
	\end{equation}
	for $j\in\{1,\ldots,N\}$.
	Therein, the $j$th subsystem Hamiltonian $\hat H_{a_1,\ldots,a_{j-1},a_{j+1},\ldots,a_N}$ depends on the state of the other parties at a given time, with
	\begin{equation}
		\label{eq:PartiallyReducedHamiltonian}
	\begin{aligned}
		&\hat H_{a_1,\ldots,a_{j-1},a_{j+1},\ldots,a_N}
		\\
		=&\langle a_1,\ldots,a_{j-1}|{\otimes}\hat 1_j{\otimes}\langle a_{j+1},\ldots,a_N|
		\hat H
		\\
		&\times| a_1,\ldots,a_{j-1}\rangle{\otimes}\hat 1_j{\otimes}| a_{j+1},\ldots,a_N\rangle.
	\end{aligned}
	\end{equation}
	Thereby, the evolution of each subsystem's pure state is forced to remain in tensor-product form, while still depending of the current states of all other parties.

	Analogously to the actual von~Neumann equation, the derivation of the equation of motion \eqref{eq:SepvNeumann} for local states is deduced from applying the principle of least action, with the constraint of factorizability \cite{SW20}.
	Please note that the Hamiltonian $\hat H$ used is the same as in the unconstrained von~Neumann equation.
	Because of the coupling induced by $\hat H_{a_1,\ldots,a_{j-1},a_{j+1},\ldots,a_N}$, the subsystems are still interacting while not being able to produce entanglement \cite{SW17,SW20}.
	Now, the idea is that we can compare the speed of the system with the maximal speed under the non-entangling dynamics to assess entanglement-caused quantum speedups.

\begin{widetext}

\subsubsection{Derivation.}

	From Eq. \eqref{eq:SepvNeumann} and the product rule of differentiation, we can formulate the evolution of the composite state,
	\begin{equation}
		\label{eq:SensitySepEoM}
	\begin{aligned}
		&\partial_t\left(|a_1,\ldots,a_N\rangle\langle a_1,\ldots,a_N|\right)
		\\
		=&\sum_{j=1}^N
		| a_1,\ldots,a_{j-1}\rangle\langle a_1,\ldots,a_{j-1}|
		\otimes\frac{1}{i\hbar}\big[\hat H_{a_1,\ldots,a_{j-1},a_{j+1},\ldots,a_N},|a_j\rangle\langle a_j|\big]
		\otimes
		| a_{j+1},\ldots,a_N\rangle\langle a_{j+1},\ldots,a_N|
		\\
		=&\frac{1}{i\hbar}\left[
			\sum_{j=1}^N
			\hat 1_{1}\otimes\cdots\otimes \hat 1_{j-1}
			\otimes\hat H_{a_1,\ldots,a_{j-1},a_{j+1},\ldots,a_N}\otimes
			\hat 1_{j+1}\otimes\cdots\otimes \hat 1_{N}
			,|a_1,\ldots,a_N\rangle\langle a_1,\ldots,a_N|
		\right],
	\end{aligned}
	\end{equation}
	using the partially reduced operators in Eq. \eqref{eq:PartiallyReducedHamiltonian}.
	In the next steps, we proceed analogously to Sec. \ref{subsec:QSLsingle} but with a generator of the dynamics that is defined via $\sum_{j=1}^N
	\hat 1_{1}\otimes\cdots\otimes \hat 1_{j-1}
	\otimes\hat H_{a_1,\ldots,a_{j-1},a_{j+1},\ldots,a_N}\otimes
	\hat 1_{j+1}\otimes\cdots\otimes \hat 1_{N}$.
	As explained in the previous single-party study, we can compute the trace norm of this commutator.
	Firstly, this yields the variance, similarly expressed via
	\begin{equation}
	\begin{aligned}
		(\gamma\hbar)^2
		=&\langle a_1,\ldots,a_N|
		\left(\Delta
			\sum_{j=1}^N
			\hat 1_{1}\otimes\cdots\otimes \hat 1_{j-1}
			\otimes\hat H_{a_1,\ldots,a_{j-1},a_{j+1},\ldots,a_N}\otimes
			\hat 1_{j+1}\otimes\cdots\otimes \hat 1_{N}
		\right)^2
		|a_1,\ldots,a_N\rangle
		\\
		=&
		\sum_{j=1}^N\langle a_j|\left(\hat H_{a_1,\ldots,a_{j-1},a_{j+1},\ldots,a_N}\right)^2|a_j\rangle
		+\sum_{\substack{j,j'=0 \\ j\neq j'}}^N
		\langle a_j|\hat H_{a_1,\ldots,a_{j-1},a_{j+1},\ldots,a_N}|a_j\rangle
		\langle a_{j'}|\hat H_{a_1,\ldots,a_{j'-1},a_{j'+1},\ldots,a_N}|a_{j'}\rangle
		-E^2
		\\
		=&\sum_{j=1}^N\langle a_j|(\Delta\hat H_{a_1,\ldots,a_{j-1},a_{j+1},\ldots,a_N})^2|a_j\rangle
		,
	\end{aligned}
	\end{equation}
	with $E=\sum_{j=1}^N\langle a_j|\hat H_{a_1,\ldots,a_{j-1},a_{j+1},\ldots,a_N}|a_j\rangle$ and a similar notation as used above, i.e., $\Delta\hat H_{a_1,\ldots,a_{j-1},a_{j+1},\ldots,a_N}=\hat H_{a_1,\ldots,a_{j-1},a_{j+1},\ldots,a_N}-\langle a_j|\hat H_{a_1,\ldots,a_{j-1},a_{j+1},\ldots,a_N}|a_j\rangle$.
	In the form above, $(\hbar\gamma)^2$ is the sum of the subsystems' energy variances in a separable configuration.
	Our previous considerations from Sec. \ref{subsec:QSLsingle} tell us that, for each $j$ in the sum, this variance is optimized via the difference of the minimal and maximal eigenvalue of the corresponding operator $\hat H_{a_1,\ldots,a_{j-1},a_{j+1},\ldots,a_N}$.
	However, these eigenvalues depend on the indexed states of this partially reduced Hamiltonian.
	To mitigate that, we can determine the energy bounds of all $\hat H_{a_1,\ldots,a_{j-1},a_{j+1},\ldots,a_N}$ by employing the inequality
	\begin{equation}
		\label{eq:supIneq}
		\sup_{|a_j\rangle}\,
		\langle a_j| \hat H_{a_1,\ldots,a_{j-1},a_{j+1},\ldots,a_N}|
		a_j \rangle
		\leq \sup_{|a_1,\ldots,a_N\rangle}
		\langle a_1,\ldots,a_N|\hat H|a_1,\ldots,a_N\rangle,
	\end{equation}
	for the supremum and the analogous inequality for the infimum, which applies to each of the $N$ $j$s.
	Combining the thereby estimated minimal and maximal energy values to express the maximized variance (see Sec. \ref{subsec:QSLsingle}), we then find that the separable quantum speed limit is bounded by
	\begin{equation}
		\label{eq:sepQSLintermediate}
		\mathsf{QSL}_\mathrm{sep}\leq \left[N\,
		\left(\frac{
			\sup_{|a_1,\ldots,a_N\rangle}
			\langle a_1,\ldots,a_N|\hat H|a_1,\ldots,a_N\rangle
			-\inf_{|a_1,\ldots,a_N\rangle}
			\langle a_1,\ldots,a_N|\hat H|a_1,\ldots,a_N\rangle
		}{\hbar}\right)^2\right]^{1/2}.
	\end{equation}

	The previous formula implies that the last task for limiting the rate of change of a composite system when confined to separable states is finding the maximal and minimal expectation value of $\hat H$ for product states, similarly to the single system in Sec. \ref{subsec:QSLsingle}.
	Such an optimization task over product states has been carried out in Ref. \cite{SV13}.
	This led to so-called separability eigenvalue equations,
	\begin{equation}
		\label{eq:SepEvalEqs}
		\hat H_{a_1,\ldots,a_{j-1},a_{j+1},\ldots,a_N}|a_j\rangle
		=E_\mathrm{sep}|a_j\rangle,
		\quad\text{for}\quad
		j\in\{1,\ldots,N\},
	\end{equation}
	using the notation introduced in Eq. \eqref{eq:PartiallyReducedHamiltonian} for $\hat H_{a_1,\ldots,a_{j-1},a_{j+1},\ldots,a_N}$.
	In the set of coupled eigenvalue equations, Eq. \eqref{eq:SepEvalEqs}, $E_\mathrm{sep}$ is dubbed separability eigenvalue of $\hat H$, and the state $|a_1,\ldots,a_N\rangle$ refers to as a separability eigenvector.
	With this equation, the sought-after maximal and minimal energies for product states are respectively determined by the maximal and minimal separability eigenvalue,
	\begin{equation}
		E_\mathrm{\max\,sep}=\sup\{E_\mathrm{sep}:E_\mathrm{sep}\text{ solves Eq. \eqref{eq:SepEvalEqs}}\}
		\quad\text{and}\quad
		E_\mathrm{\min\,sep}=\inf\{E_\mathrm{sep}:E_\mathrm{sep}\text{ solves Eq. \eqref{eq:SepEvalEqs}}\},
	\end{equation}
	which can be inserted into Eq. \eqref{eq:sepQSLintermediate}.
	Thus, in conclusion, the propagation of processes confined to the separable quantum states for $N$ parties is bounded by
	\begin{equation}
		\label{eq:QSLsep}
		\mathsf{QSL}_\mathrm{sep}
		\leq
		\sqrt N\,\frac{E_\mathrm{\max\,sep}-E_\mathrm{\min\,sep}}{\hbar}
		=\mathsf{QSL}_\mathrm{sep}^{+}.
	\end{equation}
\end{widetext}

\subsection{Preliminary discussion}

	We derived a local quantum speed limit by bounding the rate of change of separable states, which was based on the corresponding equations of motions that confine the evolution to the manifold of separable states \cite{SW17,SW20}.
	Analogously to the single-party scenario, we found that the bound $\mathsf{QSL}_\mathrm{sep}^+$ [Eq. \eqref{eq:QSLsep}] can be expressed in terms of the difference of maximal and minimal expectation value for product states, i.e., the maximal and minimal separability eigenvalue.
	These extreme values are obtained by solving a set of coupled eigenvalue equations \cite{SV13} of the system's Hamiltonian, Eq. \eqref{eq:SepEvalEqs}, including arbitrary interactions between the $N$ parties.

	The approach of so-called separability Schr\"odinger equations, and its generalization to separability von~Neumann equations [Eq. \eqref{eq:SepvNeumann}], was used previously to distinguish separable and inseparable quantum trajectories \cite{SW17}.
	For example, the evolution of a macroscopic ensemble of pairwise interacting quantum particles was analyzed \cite{SW20}, showing distinctively different dynamical signatures in time-scales and oscillatory amplitudes for inseparable and separable dynamics.

	The methodology of separability eigenvalue equations [Eq. \eqref{eq:SepEvalEqs}] was harnessed to explore---in theory and experiments---entanglement in various physical systems with complex multipartite quantum correlations \cite{PFSV13,GSVCRTF15,GSVCRTF16,SW17macro}.
	Also, this generalized eigenvalue problem allows one to optimally decompose entangled states in terms of quasiprobabilities \cite{SW18,SMBBS19}.

	The here-derived inequality \eqref{eq:QSLsep} renders it possible to determine a limit for the rate of change of a process when confined to the manifold of separable states.
	If an observed rate of change exceeds $\mathsf{QSL}_\mathrm{sep}^+$, we can confirm an entanglement-assisted speedup of the system's evolution.
	Moreover, since the unconstrained speed limit $\mathsf{QSL}$ [Eq. \eqref{eq:QSL}] is a tight bound in finite-dimensional systems, we find that states can exceed the speed limit $\mathsf{QSL}_\mathrm{sep}^+$ when $\mathsf{QSL}>\mathsf{QSL}_\mathrm{sep}^+$ holds true.
	Furthermore, for the trivial case of a single party, $N=1$, $\mathsf{QSL}_\mathrm{sep}^+=\mathsf{QSL}$ holds true since the single-party separability eigenvalue equations are just eigenvalue equations, providing a consistency check.

	In the remainder of this work, we study the properties of the separable quantum speed limit and apply it to concrete examples.
	Specifically, we begin with interesting applications in Sec. \ref{sec:application},
	and we provide general properties and generalization of our method in Sec. \ref{sec:PropertiesGeneralizations}.


\section{Applications}
\label{sec:application}

	Now, insightful examples are provided to demonstrate the entanglement-assisted quantum speedup by quantifying to which extent separable speed limits are overcome by entangling processes.
	The first demonstration concerns a spin-spin coupling, Sec. \ref{subsec:TwoQubits}, being an example to realize a swap quantum gate.
	The second application, in Sec. \ref{subsec:TwoQudits}, pertains to two high-dimensional interacting systems.
	Finally, our third example concerns highly multipartite entanglement for optical qubits, Sec. \ref{subsec:MultipleQubits}.
	Importantly, the last application exhibits an exponential speedup with the number of interacting parties.

	In our earlier derivation, we determined the exact quantum speed limit $\mathsf{QSL}$ [Eq. \eqref{eq:QSL}] and estimated the separable speed limit through $\mathsf{QSL}_\mathrm{sep}\leq\mathsf{QSL}_\mathrm{sep}^+$ [Eq. \eqref{eq:QSLsep}].
	In addition, we can lower-bound the entanglement-assisted quantum speedup via the ratio
	\begin{equation}
		\label{eq:RatioSpeedup}
		\frac{\mathsf{QSL}}{\mathsf{QSL}_\mathrm{sep}}\geq
		\frac{\mathsf{QSL}}{\mathsf{QSL}^+_\mathrm{sep}}.
	\end{equation}
	The right-hand side of the inequality yields a useful quantifier for probing by which margin the separable speed limit is (at least) beaten by different physical processes.

\subsection{Two interacting qubits}
\label{subsec:TwoQubits}

	As an essential proof of concept, we explore two qubits that interact via a spin-spin coupling.
	Already taking a result from the next section (Sec. \ref{sec:PropertiesGeneralizations}) into account, we can focus on the interaction Hamiltonian
	\begin{equation}
		\label{eq:twoqubitHamiltonian}
		\hat H=\frac{\hbar\kappa}{2}\left(
			\hat\sigma_x\otimes\hat\sigma_x
			+\hat\sigma_y\otimes\hat\sigma_y
			+\hat\sigma_z\otimes\hat\sigma_z
		\right)
	\end{equation}
	alone and ignore the local evolution.
	The operators $\hat\sigma_{w}$ for $w\in\{ x,y,z\}$ denote Pauli-spin matrices, and $\kappa\in\mathbb R\setminus\{0\}$ is the coupling strength.
	In addition, $\hat\sigma_0$ defines the qubit identity.

	Please note that the Hamiltonian under study can be expressed in terms of the swap operator $\hat V:|a,b\rangle\mapsto|b,a\rangle$ through $\hat H=\hbar\kappa\hat V-(\hbar\kappa/2)\hat\sigma_0^{\otimes 2}$ because of $2\hat V=\hat\sigma_0^{\otimes 2}+\hat\sigma_x^{\otimes 2}+\hat\sigma_y^{\otimes 2}+\hat\sigma_z^{\otimes 2}$.
	As identities will commute with all density matrices in all von~Neumann-type equations, they do not contribute to speed limits.
	Furthermore, all technical calculations for obtaining the exact results for both the separable and inseparable case for this Hamiltonian can be found in Appendix \ref{app:bipartite} and are used in the following.

	The maximal and minimal eigenvalue of $\hat H$ are
	\begin{equation}
		E_{\max}=\hbar|\kappa|-\frac{\hbar\kappa}{2}
		\text{ and }
		E_{\min}=-\hbar|\kappa|-\frac{\hbar\kappa}{2}.
	\end{equation}
	According to Eq. \eqref{eq:QSL}, this yields an upper bound to the rate of change that reads
	\begin{equation}
		\mathsf{QSL}=2|\kappa|.
	\end{equation}
	The solution of the separability eigenvalue equations \eqref{eq:SepEvalEqs} render it possible to determine the maximal and minimal separability eigenvalues,
	\begin{equation}
		E_\mathrm{max\,sep}=\hbar\kappa-\frac{\hbar\kappa}{2}
		\text{ and }
		E_\mathrm{min\,sep}=0-\frac{\hbar\kappa}{2}
	\end{equation}
	when $\kappa>0$, as well as $E_\mathrm{max\,sep}=0-\hbar\kappa/2$ and $E_\mathrm{min\,sep}=\hbar\kappa-\hbar\kappa/2$ for $\kappa<0$.
	Using Eq. \eqref{eq:QSLsep}, we thus find
	\begin{equation}
		\mathsf{QSL}_\mathrm{sep}^+=\sqrt2|\kappa|,
	\end{equation}
	for both signs of $\kappa$ and with $N=2$ involved parties.
	The ratio that allows us to determine the quantum speedup [Eq. \eqref{eq:RatioSpeedup}] is
	\begin{equation}
		\label{eq:QSLtwoqubits}
		\frac{\mathsf{QSL}}{\mathsf{QSL}_\mathrm{sep}^+}
		=\sqrt2\approx 1.414>1.
	\end{equation}
	This corresponds to an achievable speedup of $\approx 41.4\%$ due to generation of entanglement during the spin-spin interaction.

	Even more concretely, we can consider the solutions of the separable [Eq. \eqref{eq:SepvNeumann}] and inseparable [Eq. \eqref{eq:vonNeumann}] equations of motion.
	For a product state $|a_0,b_0\rangle$ at time $t=0$ (with $\langle a_0|a_0\rangle=1=\langle b_0|b_0\rangle$), we obtain the results
	\begin{equation}
		|\psi(t)\rangle
		=\cos(\kappa t)|a_0,b_0\rangle
		-i\sin(\kappa t)|b_0,a_0\rangle
	\end{equation}
	for the entangling evolution and
	\begin{equation}
		|\psi_\mathrm{sep}(t)\rangle=|a(t)\rangle\otimes|b(t)\rangle
	\end{equation}
	for the separable scenario, where
	\begin{equation}
	\begin{aligned}
		|a(t)\rangle
		=&\cos(|q|\kappa t)|a_0\rangle
		-i\frac{q^\ast}{|q|}\sin(|q|\kappa t)|b_0\rangle
		\\
		\text{and }
		|b(t)\rangle
		=&\cos(|q|\kappa t)|b_0\rangle
		-i\frac{q}{|q|}\sin(|q|\kappa t)|a_0\rangle
		,
	\end{aligned}
	\end{equation}
	with
	\begin{equation}
		\label{eq:qubitoverlap}
		q=\langle a_0|b_0\rangle.
	\end{equation}
	See Appendix \ref{app:bipartite} (and Ref. \cite{SW17}) for details.
	We emphasize that both local states $|a(t)\rangle$ and $|b(t)\rangle$ depend on both initial states $|a_0\rangle$ and $|b_0\rangle$.
	This shows that the qubits interact across local systems although the evolution is confined to product states.

	With those results, we can, for example, determine the time it takes to swap local states.
	That is, we assess the physical time it takes to implement a swap gate, rather than the number of quantum-computational steps to achieve this, which is the common quantum-information-based approach.
	The full swap happens in both cases after a quarter period.
	At those times, we obtain a separable state, $|\psi(T/4)\rangle=|b_0,a_0\rangle=|\psi_\mathrm{sep}(T_\mathrm{sep}/4)\rangle$, that exchanges the initial qubits, with
	\begin{equation}
		T=\frac{2\pi}{|\kappa|}
		\quad\text{and}\quad
		T_\mathrm{sep}=\frac{2\pi}{|q|\,|\kappa|}.
	\end{equation}
	For $0< |q|\leq 1$, this means that executing the swap gate takes a factor of $1/|q|$ less time for the actual evolution when compared with the non-entangling process, despite the final outcome being separable in either case.

\begin{figure}
	\includegraphics[width=\columnwidth]{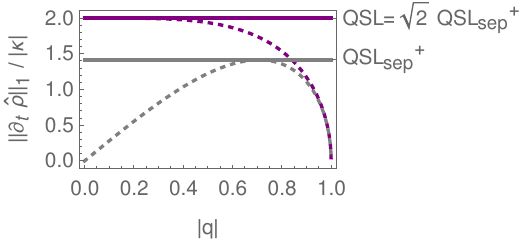}
	\caption{%
		Entanglement-assisted speedup as a function of the two-qubit overlap $|q|$, Eq. \eqref{eq:qubitoverlap}.
		Dashed curves show the rate of change (scaled to the coupling parameter $|\kappa|$) for the inseparable [upper curve, Eq. \eqref{eq:speedTQ}] and separable [lower curve, Eq. \eqref{eq:speedTQsep}] dynamics.
		The process under study is a spin-spin coupling for realizing a swap gate; see Eq. \eqref{eq:twoqubitHamiltonian} for the Hamiltonian.
		Solid horizontal lines indicate ultimate speed limits for the separable and inseparable evolution.
		The overhead in speed in relation to the separable speed limit is highest for $|q|\to0$, i.e., orthogonal initial local states.
		The entangling process beats the separable speed limit $\mathsf{QSL}_\mathrm{sep}^+$ for all $|q|$ below $2^{-1/4}\approx0.84$.
		For almost identical initial states, $|q|\approx 1$, the entanglement-based quantum advantage disappears.
	}\label{fig:twoqubits}
\end{figure}

	Furthermore, we can compute the trace norms of the exact rates of changes.
	We readily find
	\begin{subequations}
	\begin{align}
		\label{eq:speedTQ}
		\left\|
			\partial_t |\psi(t)\rangle\langle\psi(t)|
		\right\|_1
		=&2|\kappa|\sqrt{1-|q|^4}
		\quad\text{and}
		\\
		\label{eq:speedTQsep}
		\left\|
			\partial_t |\psi_\mathrm{sep}(t)\rangle\langle\psi_\mathrm{sep}(t)|
		\right\|_1
		=&2\sqrt{2}|q|\,|\kappa|\sqrt{1-|q|^2}.
	\end{align}
	\end{subequations}
	Thereby, we also find that the separable bound is tight here since we have $2\sqrt{2}|q|\,|\kappa|\sqrt{1-|q|^2}=\sqrt{2}|\kappa|=\mathsf{QSL}_\mathrm{sep}^+$ for $|q|^2=1/2$;
	see also the definition \eqref{eq:qubitoverlap}.
	The overall entanglement-induced speedup for the process under study is depicted in Fig. \ref{fig:twoqubits} for all possible initial conditions.

	In addition to this proof of concept example for two qubits that demonstrate an entanglement-caused speedup for a swap-gate operation, the following two applications of our methodology answer the vital question of how the entanglement-induced speedup scales with the local dimension of subsystems, as well as the number of interacting parties.

\subsection{High-dimensional bipartite systems}
\label{subsec:TwoQudits}

	For the second application, suppose two interacting $d$-level quantum systems, qudits.
	The ground state $|\psi_0\rangle$ is assumed to be energetically well-separated from all other eigenstates of the Hamiltonian.
	Thus, the latter collectively define an orthogonal space to the ground state, represented through the projector $\hat 1_1\otimes\hat 1_2-|\psi_0\rangle\langle\psi_0|$, with local identities $\hat 1_j$ for $j=1,2$.
	The energies of the ground state and its orthogonal complement are $E_0$ and $E_\perp$, respectively, with $E_0\ll E_\perp$.
	Hence, the system's dynamics is approximated by the Hamiltonian
	\begin{equation}
	\begin{aligned}
		\label{eq:HamiltonianHD}
		\hat H
		=&E_0|\psi_0\rangle\langle\psi_0|
		+E_\perp\Big(
			\hat 1_1\otimes\hat 1_2-|\psi_0\rangle\langle\psi_0|
		\Big)
		\\
		=&-(E_\perp-E_0)|\psi_0\rangle\langle\psi_0| +E_\perp\hat 1_1\otimes\hat 1_2.
	\end{aligned}
	\end{equation}
	Again, all exact results that we apply in the following can be found in Appendix \ref{app:bipartite}.

	For determining $\mathsf{QSL}$, the minimal and maximal eigenvalues can be directly obtained,
	\begin{equation}
		E_{\min}=E_0
		\quad\text{and}\quad
		E_{\max}=E_\perp.
	\end{equation}
	Now suppose the ground state has a Schmidt decomposition
	\begin{equation}
		\label{eq:maxqudit}
		|\psi_0\rangle=\frac{1}{\sqrt{d}}\sum_{n=0}^{d-1}|n\rangle\otimes|n\rangle,
	\end{equation}
	being a maximally entangled state.
	Using the solutions of separability eigenvalue equations \eqref{eq:SepEvalEqs} for the rank-one operator plus identity in Eq. \eqref{eq:HamiltonianHD} (see also Appendix \ref{app:bipartite} and Refs. \cite{SV09,SV13}), the extreme separability eigenvalues read
	\begin{equation}
	\begin{aligned}
		E_\mathrm{\min\,sep}=&-(E_\perp-E_0)\frac{1}{d}+E_\perp
		\quad\text{and}\\
		E_\mathrm{\max\,sep}=&-(E_\perp-E_0)\cdot0+E_\perp=E_\perp.
	\end{aligned}
	\end{equation}

\begin{figure}
	\includegraphics[width=\columnwidth]{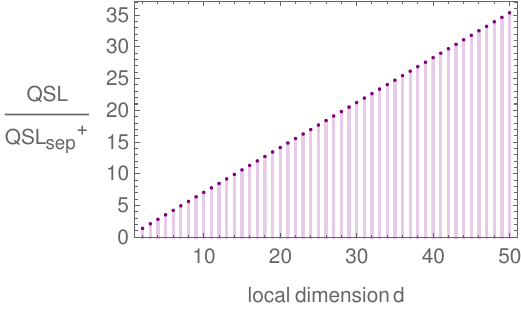}
	\caption{%
		Maximal speedup $\mathsf{QSL}/\mathsf{QSL}_\mathrm{sep}^+$ of the process described through the Hamiltonian in Eq. \eqref{eq:HamiltonianHD} whose ground state is a maximally entangled qudit state [Eq. \eqref{eq:maxqudit}] for $d$-dimensional local spaces.
		Like in Fig. \ref{fig:twoqubits}, we find an increased speed by a factor $\sqrt 2$ for qubits, $d=2$.
		From there, the impact of the entangling dynamics increases proportionally with the local dimension $d$, as expressed in Eq. \eqref{eq:SpeedupHD}.
		For instance, already the $d=3$ yields a ratio $3/\sqrt2\approx2.12$, i.e., a $112\%$ speed increase compared to the fastest possible separable evolution for the given process.
	}\label{fig:highdimensional}
\end{figure}

	Applying our previous results, Eqs. \eqref{eq:QSL} and \eqref{eq:QSLsep}, provides the sought-after speed limits
	\begin{equation}
		\mathsf{QSL}=\frac{E_\perp-E_0}{\hbar}
		\quad\text{and}\quad
		\mathsf{QSL}^+_\mathrm{sep}=\sqrt{2}\frac{E_\perp-E_0}{\hbar d}.
	\end{equation}
	Thus, the speedup as described through the ratio in Eq. \eqref{eq:RatioSpeedup} can be expressed as
	\begin{equation}
		\label{eq:SpeedupHD}
		\frac{\mathsf{QSL}}{\mathsf{QSL}^+_\mathrm{sep}}=\frac{d}{\sqrt2}.
	\end{equation}
	This ratio exceeds one for all $d\geq2$ and increases linearly with the subsystem's dimension;
	see Fig. \ref{fig:highdimensional}.
	Thus, the higher the dimension of the composite system, whose ground state is a maximally entangled state, the higher the entanglement-assisted speedup can be.

\subsection{Highly multipartite systems}
\label{subsec:MultipleQubits}

	For this last application, being arguably the most intriguing one,  we study an $N$-mode nonlinear optical process.
	Specifically, the interaction Hamiltonian is
	\begin{equation}
		\label{eq:HamiltonianNparty}
		\hat H=\hbar\kappa\left(\hat a^\dag\right)^{\otimes K}\otimes\hat a^{\hat\otimes (N-K)}+\hbar\kappa^\ast\hat a^{\otimes K}\otimes\left(\hat a^\dag\right)^{\otimes (N-K)}.
	\end{equation}
	Through this process, photons are produced via each local creation operator $\hat a^\dag$ in the first $K$ modes on the expense of the annihilation ($\hat a$) of photons in the remaining modes, and vice versa.
	The complex parameter $\kappa\in\mathbb C\setminus\{0\}$ describes the interaction strength.

	Typically, processes with a high nonlinearity are quite weak and at most one photon is produced (or absorbed) per mode.
	Hence, we approximate each local Hilbert space via $|0\rangle$ (vacuum) and $|1\rangle$ (single photon).
	Then, each local annihilation operator can be represented as the following qubit operator:
	\begin{equation}
		\hat a|1\rangle=|0\rangle
		\quad\text{and}\quad
		\hat a|0\rangle=0
		\quad\Rightarrow\quad
		\hat a=|0\rangle\langle 1|.
	\end{equation}
	And Hermitian conjugation leads to $\hat a^\dag=|1\rangle\langle 0|$ for the creation operator in the local Fock space of at most one photon.
	Avoiding approximations, we can alternatively think of the system as a spin system, where $|0\rangle=|\downarrow\rangle$ and $|1\rangle=|\uparrow\rangle$, and in terms of fermionic field operators $\hat a$, allowing for one excitation at most because of the exclusion principle.

	As before, we moved the exact, technical calculations to the appendix (Appendix \ref{app:multipartite}),
	and, in the following, we focus on discussing the physical implications.

	The minimal and maximal eigenvalues of $\hat H$ in Eq. \eqref{eq:HamiltonianNparty} are
	\begin{equation}
		E_{\max}=\hbar|\kappa|
		\quad\text{and}\quad
		E_{\min}=-\hbar|\kappa|.
	\end{equation}
	This means the maximal speed of the quantum process is
	\begin{equation}
		\mathsf{QSL}=\frac{E_{\max}-E_{\min}}{\hbar}=2|\kappa|.
	\end{equation}
	For instance, GHZ-like states show this particular rate of change (Appendix \ref{app:multipartite}).

	By contrast, for $N$-separable states, we solve the $N$-partite separability eigenvalue equations of $\hat H$ and obtain
	\begin{equation}
		E_\mathrm{max\,sep}=\frac{\hbar|\kappa|}{2^{N-1}}
		\quad\text{and}\quad
		E_\mathrm{min\,sep}=-\frac{\hbar|\kappa|}{2^{N-1}}.
	\end{equation}
	Therefore, the $N$-separable speed limit is determined by
	\begin{equation}
		\mathsf{QSL}_\mathrm{sep}^+
		=\sqrt{N}\frac{E_\mathrm{max\,sep}-E_\mathrm{min\,sep}}{\hbar}=\sqrt{N}\frac{|\kappa|}{2^{N-2}}.
	\end{equation}
	Consequently, the ratio in Eq. \eqref{eq:RatioSpeedup} to assess the quantum speedup caused by entanglement reads
	\begin{equation}
		\label{eq:SpeedupMulti}
		\frac{\mathsf{QSL}}{\mathsf{QSL}_\mathrm{sep}^+}=\frac{2^{N-1}}{\sqrt{N}}.
	\end{equation}

\begin{figure}
	\includegraphics[width=\columnwidth]{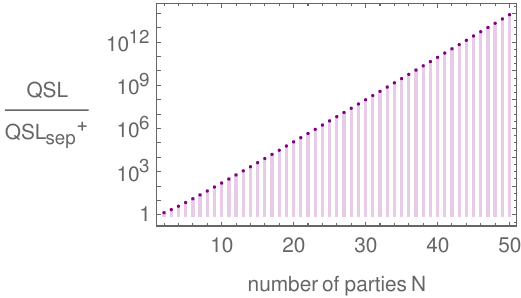}
	\caption{%
		Entanglement-assisted quantum speedup on a logarithmic scale for an $N$-mode nonlinear optical process, Eq. \eqref{eq:HamiltonianNparty}.
		The improvement of the speed of the entangling evolution is exponential, Eq. \eqref{eq:SpeedupMulti}, when compared with the same dynamics that is confined to the manifold of separable states.
		For two qubits, $N=2$, we obtain the same ratio $\sqrt{2}$ as in Fig. \ref{fig:twoqubits}.
		However, already for a tripartite and fourpartite system of qubits, we have a ratio of $4/\sqrt{3}\approx 2.3$ and $4$, respectively, a $130\%$ and $300\%$ increase in speed.
	}\label{fig:multimode}
\end{figure}

	The process under study operates exponentially faster than one can ever expect from the separable evolution; cf. Fig. \ref{fig:multimode}.
	This holds true independently of the coupling constant $\kappa$, assuming $|\kappa|\neq0$.
	Recall that this exponential gain is a physical reasoning for a quantum speedup and not necessarily connected with quantum-complexity-based concepts of exponential quantum performance due to entanglement.
	Still, one would naturally expect that the complexity-based and physical speedup are related.
	The specific relation, however, remains unknown to date and opens up a research direction for future investigations.


\section{Generalizations}
\label{sec:PropertiesGeneralizations}

	On top of the discussed application, we now study general properties that can be concluded from our findings.
	Firstly, in Sec. \ref{subsec:witnesses}, we formulate experimentally accessible criteria through the separable speed limit for quantum correlations between non-infinitesimally separated times.
	Secondly, it is shown that free processes---describing purely local contributions to the dynamics in the context of entanglement resources---do not play a role when characterizing the difference of speed limits, Sec. \ref{subsec:interactionpicture}.
	Eventually, estimates of local quantum speed limits for open systems are provided in Sec. \ref{subsec:opensystems}.

\subsection{Non-infinitesimal changes for measurable criteria}
\label{subsec:witnesses}

	Thus far, we studied infinitesimal rates of change, i.e., $\|\partial_t\hat\rho(t)\|_1$.
	Particularly, we found that the speed limit of local processes is bounded as described in Eq. \eqref{eq:QSLsep}.
	In other words, when the evolution is confined to the domain of separable states, we have
	\begin{equation}
		\label{eq:SepInfinChange}
		\|\partial_t\hat\rho_\mathrm{sep}(t)\|_1\leq \mathsf{QSL}^+_\mathrm{sep}.
	\end{equation}

	Now, we consider a finite time interval, $0\leq t\leq T$, with the initial time $0$ and the final time $T$.
	We can formally always write $\hat\rho_\mathrm{sep}(T)=\hat\rho_\mathrm{sep}(0)+\int^{T}_{0}dt\,\partial_t\hat\rho_\mathrm{sep}(t)$.
	Applying this relation, together with the triangle inequality and the bound in Eq. \eqref{eq:SepInfinChange}, we can write
	\begin{equation}
	\begin{aligned}
		\|\hat\rho_\mathrm{sep}(T)-\hat\rho_\mathrm{sep}(0)\|_{1}
		\leq&\int_{0}^{T}dt\,\|\hat\rho_\mathrm{sep}(t)\|
		\\
		\leq&\int_{0}^{T}dt\,\mathsf{QSL}^+_\mathrm{sep}		=T\cdot\mathsf{QSL}^+_\mathrm{sep}.
	\end{aligned}
	\end{equation}
	This inequality upper-bounds the change in the finite time interval $[0,T]$.
	Conversely, a violation of this bound is inconsistent with the process operating in a separable manner, implying the utilization of entanglement in the dynamics that carries the system from the initial state to the final state at time $T$.
	For convenience, the above relation can be recast into the form of a difference quotient also,
	\begin{equation}
		\left\|
			\frac{\hat\rho_\mathrm{sep}(T)-\hat\rho_\mathrm{sep}(0)}{T-0}
		\right\|_{1}\leq \mathsf{QSL}^+_\mathrm{sep}.
	\end{equation}

	The latter form is particularly useful when determining changes in measured quantities.
	Let $\hat L=\hat L^\dag$ be an observable.
	Then the H\"older inequality---here in the form $|\mathrm{tr}(\hat\epsilon\hat L)|\leq\|\hat\epsilon\|_1\cdot\|\hat L\|_\infty$, using the spectral norm $\|\hat L\|_\infty$ that is defined through the maximum of the absolute eigenvalues of $\hat L$---can be used to provide a bound for the changes of expectation values.
	Namely, we have
	\begin{equation}
		\left|
			\mathrm{tr}\left(
				\frac{\hat\rho_\mathrm{sep}(t_f)-\hat\rho_\mathrm{sep}(t_i)}{t_f-t_i}
				\hat L
			\right)
		\right|\leq \mathsf{QSL}^+_\mathrm{sep}\|\hat L\|_\infty,
	\end{equation}
	applying the previous estimate of finite changes and using a straightforward generalization to arbitrary intervals, $t\in[t_i,t_f]$.
	Equivalently, we can say that the change of the expectation value $\langle \hat L\rangle_t=\mathrm{tr}(\hat\rho_\mathrm{sep}(t)\hat L)$ is bounded by
	\begin{equation}
		\label{eq:MeasurableBound}
		\left|\langle \hat L\rangle_{t_f}-\langle \hat L\rangle_{t_i}\right|
		\leq \big(t_f-t_i\big)\mathsf{QSL}^+_\mathrm{sep}
		\|\hat L\|_\infty
	\end{equation}
	for the separable evolution.
	A violation of this temporal (specifically, two-time) separability constraint provides a measurable means to certify an entangling evolution.

	In a sense, the here-derived bound for expectation values bears some resemblance to notions that are captured by temporal correlations via the Leggett--Garg inequality \cite{LG85} and nonlocal properties of Clauser--Horne--Shimony--Hold inequality \cite{CHSH69} for two measurement settings.
	Also, the statement in Eq. \eqref{eq:MeasurableBound} can be interpreted as
	\begin{equation}
		\label{eq:boundingCone}
		\langle\hat L\rangle_0
		-t\,\mathsf{QSL}^+_\mathrm{sep}\,\|\hat L\|_\infty
		\leq
		\langle\hat L\rangle_t
		\leq
		\langle\hat L\rangle_0
		+t\,\mathsf{QSL}^+_\mathrm{sep}\,\|\hat L\|_\infty
		.
	\end{equation}
	This means, the functional $\langle\hat L\rangle_t$ is bounded to a two-dimensional cone in a $t$-$\langle \hat L\rangle_t$ diagram for separable dynamics.
	Leaving this cone at any time $t$ thus shows an entanglement-enhanced evolution.
	In addition, it is noteworthy that $\langle\hat L\rangle_t$ in Eq. \eqref{eq:boundingCone} can serve as an entanglement witness, e.g., via the construction in Ref. \cite{SV13}, for a state at time $t$ that is entangled and thus not attainable through a non-entangling dynamics.

	Very recently, several works considered speed limits in the context of observables and entanglement.
	For instance, the speed limit of observables akin to our two-point correlators were considered in Ref. \cite{MP22}.
	Furthermore, the dynamics of entanglement measures was explored in Ref. \cite{GC22,PSMDP22}, utilizing Feynman's sum over histories and correlation functions.
	In these approaches, however, the speed limit of processes confined to the manifold of separable states was not an accessible quantity.
	This hurdle is overcome with our contribution.

\subsection{Local processes and interaction picture}
\label{subsec:interactionpicture}

	An evolution that encompasses several subsystems decomposes into local contributions and interaction terms.
	For instance, an $N$-partite Hamiltonian $\hat H$ may be written as
	\begin{equation}
		\label{eq:LocAndIntHamiltonian}
		\hat H=\sum_{k=1}^N \hat 1_1\otimes\cdots\otimes\hat 1_{k-1}\otimes \hat H_k\otimes\hat 1_{k+1}\otimes\cdots\otimes\hat 1_N
		+\hat H_\mathrm{int},
	\end{equation}
	where each $\hat H_j$ describes local dynamics and $\hat H_\mathrm{int}$ includes arbitrary interactions between subsystems.
	In the following, we exploit the interaction picture to show that local contributions can be safely ignored since they contribute equally to the separable and inseparable evolution.

	Say $\hat U_j(t)$ for $j\in\{1,\ldots,N\}$ is a unitary map that satisfies $i\hbar\partial_t\hat U_j=\hat H_j\hat U_j$ and $\hat U_j(0)=\hat 1_j$.
	Furthermore, a local transformation is given by
	\begin{equation}
		\label{eq:interactionpicturedef}
		\hat\rho_\mathrm{int}=\left[
			\hat U_1\otimes\cdots\otimes\hat U_N
		\right]^\dag\hat\rho\left[
			\hat U_1\otimes\cdots\otimes\hat U_N
		\right],
	\end{equation}
	constituting the interaction picture.
	Then, the dynamics of $\hat\rho_\mathrm{int}$ [Eq. \eqref{eq:vonNeumann}] is governed by
	\begin{equation}
		\partial_t\hat\rho_\mathrm{int}=\frac{1}{i\hbar}[\hat H_\mathrm{eff},\hat\rho_\mathrm{int}],
	\end{equation}
	with $\hat H_\mathrm{eff}=\left[
		\hat U_1\otimes\cdots\otimes\hat U_N
	\right]^\dag\hat H_\mathrm{int}\left[
		\hat U_1\otimes\cdots\otimes\hat U_N
	\right]$ denoting the effective Hamiltonian in the interaction picture.
	Analogously, the separable evolution can be recast into the interaction picture since the transformation in Eq. \eqref{eq:interactionpicturedef} involves local unitaries only \cite{SW17}.
	That is, we map $|b_j\rangle=\hat U_j^\dag|a_j\rangle$ and compute
	\begin{equation}
	\begin{aligned}
		\partial_t(|b_j\rangle\langle b_j|)
		=&-\frac{1}{i\hbar}\hat U_j^\dag\big[\hat H_j,|a_j\rangle\langle a_j|\big]\hat U_j
		\\
		&+\frac{1}{i\hbar}\hat U_j^\dag\big[\hat H_{a_1,\ldots,a_{j-1},a_{j+1},\ldots,a_N},|a_j\rangle\langle a_j|\big]\hat U_j
	\end{aligned}
	\end{equation}
	for the separability equations of motion \eqref{eq:SepvNeumann}.
	For the Hamiltonian in Eq. \eqref{eq:LocAndIntHamiltonian}, the partially reduced operator takes the form
	\begin{equation}
	\begin{aligned}
		\hat H_{a_1,\ldots,a_{j-1},a_{j+1},\ldots,a_N}
		=&\sum_{k:k\neq j}\langle a_k| \hat H_k|a_k\rangle\hat 1_j
		+\hat H_j
		\\
		&+\hat U_j\left(\hat H_\mathrm{eff}\right)_{b_1,\ldots,b_{j-1},b_{j+1},\ldots,b_N}\hat U_j^\dag.
	\end{aligned}
	\end{equation}
	Therein, we applied $|a_j\rangle=\hat U_j|b_j\rangle$ and the conserved normalization, $\langle a_j|a_j\rangle=1$ for all times \cite{SW17,SW20}.
	The parts that are proportional to the identity commute with every density operator and, thus, can be ignored.
	Finally, substituting this partially reduced operator into the above equation for the derivative of $|b_j\rangle\langle b_j|$, we obtain
	\begin{equation}
		\partial_t(|b_j\rangle\langle b_j|)
		=\frac{1}{i\hbar}\big[\left(\hat H_\mathrm{eff}\right)_{b_1,\ldots,b_{j-1},b_{j+1},\ldots,b_N},|b_j\rangle\langle b_j|\big].
	\end{equation}

	In summary, the difference of the inseparable and the separable evolution is not captured by local terms, which is not surprising, still rigorously shown here.
	Speed differences between entangling and non-entangling processes are found to be due to the effective Hamiltonians in the interaction picture that carry the information about the arbitrarily complex subsystem interaction.

\subsection{Estimating speed limits for open-system dynamics}
\label{subsec:opensystems}

	Although we focus on closed systems in this work, we can extend our framework to be applicable to open quantum systems as well.
	For this purpose, take the Gorini--Kossakowski--Sudarshan--Lindblad equation,
	\begin{equation}
		\partial_t\hat\rho=\frac{1}{i\hbar}[\hat H,\hat\rho]+\mathcal D(\hat \rho)
	\end{equation}
	with the dissipative Lindblad term
	$
		\mathcal D(\hat\rho)=\sum_k (
			\hat h_k\hat\rho\hat h_k^\dag
			-\hat\rho\hat h_k^\dag\hat h_k/2
			-\hat h_k^\dag\hat h_k\hat\rho/2
		)
	$.
	Applying $\hat\rho=\sum_n p_n|\psi_n\rangle\langle\psi_n|$ and the triangle inequality, we can estimate
	\begin{equation}
	\begin{aligned}
		\left\|\partial_t\hat\rho\right\|_1
		\leq&
		\sum_n p_n
			\left\|\frac{1}{i\hbar}[\hat H,|\psi_n\rangle\langle\psi_n|]\right\|_1
		\\
		&+\sum_n p_n
			\left\|\mathcal D(|\psi_n\rangle\langle\psi_n|)\right\|_1
		\\
		\leq&\mathsf{QSL}\left(\frac{1}{i\hbar}[\hat H,\,\cdot\,]\right)
		+\mathsf{QSL}\left(\mathcal D\right),
	\end{aligned}
	\end{equation}
	where $\mathsf{QSL}(\mathcal G)$ denotes the quantum speed limit for specific generators, $\mathcal G\in\{(i\hbar)^{-1}[\hat H,\,\cdot\,],\mathcal D\}$.
	On the right-hand side of the last inequality, the first quantum speed limit is the one for the Hamiltonian $\hat H$ that too applies to closed system and has been discussed so far.
	The second contribution adds a speed that comes from the Lindblad term, given by the dissipative generator $\mathcal D$.

	Finally, we can analogously estimate the separable speed limit for the full dynamics of the open system as follows:
	\begin{equation}
		\mathsf{QSL}_\mathrm{sep}
		\leq
		\mathsf{QSL}^{+}_\mathrm{sep}\left(\frac{1}{i\hbar}[\hat H,\,\cdot\,]\right)
		+\mathsf{QSL}\left(\mathcal D\right).
	\end{equation}
	Therein, we used that the $\mathsf{QSL}\left(\mathcal D\right)$ applies to all states, including separable ones.
	(And the first term is the one from our closed-system derivation.)
	As before, a violation of this inequality shows an entanglement-assisted speedup of the open system's evolution.
	It is noteworthy that the part of the dynamics that accounts for the interaction with a bath is here upper-bounded via $\mathsf{QSL}\left(\mathcal D\right)$, thus affecting the separable and inseparable speed limit in the same manner.


\section{Conclusion}

	We formulated physical quantum speed limits for non-entangling processes.
	This consistent approach renders it possible to quantify the physical speedup of processes caused by entanglement.
	It complements the common information-based notion of quantum advantage.
	Demonstrating that a process overcomes the derived boundary certifies the entanglement-assisted speedup of the evolution.
	Several examples, including ones with arbitrarily high local dimensions as well as arbitrarily many parties, showed the fundamental influence the generation of entanglement has on quantum processes.
	Extensions of our framework are measurable criteria and generalizations to open systems.
	Our results explain why quantum entanglement enables faster computations compared to digital and analog classical computing because neither can harness entanglement as a resource.

	By confining the evolution of the system to the manifold of separable states, we derived upper bounds to the separable rate of change for arbitrary processes.
	Importantly, this constraint pertains only to states while the generator of the evolution remains unchanged.
	This still allows for interactions between subsystems.
	And any speedup beyond the derived bound is caused by additionally introducing entanglement during the propagation in time.

	Our physical approach can be related to notions of quantum advantages in information processing although the deeper relation between both notions remains unknown.
	For example, by analyzing a spin-spin correlated system, we found that the swap gate can be implemented faster when allowing to tap the resource of entanglement during the evolution, despite the outcome of the process being a separable operation that swaps two quantum registers containing a local qubit each.

	Moreover, the scaling of the entanglement-assisted quantum speedup was considered.
	We showed an exponential entanglement-driven improvement over separable speed limits as a function of the number of parties that interact in a process.
	Furthermore, we explored the possibility to increase local dimensions and its impact on the dynamical advantage.

	Beyond instantaneous rates of change, we expanded our method to apply to finite time intervals.
	This gave measurable two-time correlators to probe entanglement-assisted speedup in future experiments.
	We further proved that local contributions can be ignored when comparing speed limits although they generally produce local quantum coherence, as one expects.
	Beyond closed systems, we also generalized the methodology to open systems, leading to an additional contribution to our bounds.

	The concept of separable speed limits opens up questions for future research.
	Equations of motions were formulated that yield the propagation in time for broader kinds of manifolds, pertaining to other, more general notions of classicality \cite{SW20}.
	Therefore, different speed limits can possibly be derived to gauge the dynamic impact of other notions of quantum coherence beyond entanglement.
	Also, the general relation between the physical speedup studied here and the quantum advantage commonly considered in the context of quantum information processing---if and when established---would connect physical notions with concepts from computational complexity.
	Finally, on a technical note, we derived upper bounds of separable speeds,
	and those are tight bounds for the concrete instances we studied here.
	If these estimates can be improved in other cases, however, is currently not clear.
	This question particularly pertains to couplings to either individual or joint bath systems \cite{B02}, which shall be investigated in the future.


\begin{acknowledgments}
	The authors thank Suchitra Krishnaswamy, Yannick Freitag, and Emil Donkersloot for valuable comments and discussions.
	The authors acknowledge financial support from the Deutsche Forschungsgemeinschaft (DFG, German Research Foundation) through the Collaborative Research Center TRR~142 (Project No. 231447078, project C10).
	The work was further funded through the Ministerium f\"ur Kultur und Wissenschaft des Landes Nordrhein-Westfalen.
\end{acknowledgments}


\appendix

\section{Solutions for bipartite examples}
\label{app:bipartite}

	We consider two examples where the Hamiltonian is proportional to either $\hat V$, the swap operator, or $|\Phi\rangle\langle\Phi|$, with $|\Phi\rangle=\sum_{n=0}^{d-1}|n\rangle\otimes|n\rangle$ and $\langle\Phi|\Phi\rangle=d$.
	Here, we use an appropriately rescaled time $\tau$ to simplify equations of motions.

	The sets of extreme eigenvalues of $\hat V$ and $|\Phi\rangle\langle\Phi|$ are $\{-1,1\}$ and $\{0,d\}$, respectively.
	The corresponding sets of separability eigenvalues can be taken from Ref. \cite{SV09} and read $\{0,1\}$ in both cases.
	The unitary evolution operators that solve the von~Neumann equation take the form
	\begin{equation}
		\exp(-i\tau\hat V)=\cos(\tau)\hat 1-i\sin(\tau)\hat V
	\end{equation}
	and
	\begin{equation}
		\exp(-i\tau|\Phi\rangle\langle\Phi|)
		=
		\left(\hat 1-\frac{|\Phi\rangle\langle\Phi|}{d}\right)
		+
		\exp(-id\tau)\frac{|\Phi\rangle\langle\Phi|}{d}.
	\end{equation}
	For $|\psi(0)\rangle=|a_0,b_0\rangle$, those exponential operators lead to $|\psi(\tau)\rangle=\cos(\tau)|a_0,b_0\rangle-i\sin(\tau)|b_0,a_0\rangle$ and $|\psi(\tau)\rangle=|a_0,b_0\rangle+\langle b_0^\ast|a_0\rangle(e^{-i\tau d}-1)|\Phi\rangle/d$.

	The remainder of this appendix concerns the separable propagation in time.
	We begin with the swap operator.
	The partially reduced operators are $\hat V_a=|a\rangle\langle a|$ and $\hat V_b=|b\rangle\langle b|$.
	Thus, the separability von~Neumann equations read
	\begin{equation}
	\begin{aligned}
		i\partial_\tau|a\rangle\langle a|
		=&[|b\rangle\langle b|,|a\rangle\langle a|]
		\\
		\text{and}\quad
		i\partial_\tau|b\rangle\langle b|
		=&[|a\rangle\langle a|,|b\rangle\langle b|].
	\end{aligned}
	\end{equation}
	Comparing the right-hand sides, we find that the following operator is time-independent:
	\begin{equation}
		\hat C=|b\rangle\langle b|+|a\rangle\langle a|.
	\end{equation}
	Because of that, and using the initial states $|a_0\rangle$ and $|b_0\rangle$, we can write $\hat C=|b_0\rangle\langle b_0|+|a_0\rangle\langle a_0|$.
	Also, we can say $|b\rangle\langle b|=\hat C-|a\rangle\langle a|$, allowing us to infer $|b\rangle$ from $|a\rangle$.
	Substituting the previous decomposition into the equation of motion for $|a\rangle$, we obtain
	\begin{equation}
		\label{eq:AppAliceOnlyEoM}
		i\partial_\tau|a\rangle\langle a|=[\hat C,|a\rangle\langle a|],
	\end{equation}
	which we solve later in a more general scenario.

	The second consideration concerns $|\Phi\rangle\langle\Phi|$.
	Here, the partially reduced operators are $(|\Phi\rangle\langle\Phi|)_a=|a^\ast\rangle\langle a^\ast|$ and $(|\Phi\rangle\langle\Phi|)_b=|b^\ast\rangle\langle b^\ast|$, including the complex conjugate vectors in the computational basis.
	Analogously to the previous case, the separability von~Neumann equations can be recast as
	\begin{equation}
	\begin{aligned}
		i\partial_\tau|a\rangle\langle a|
		=&[|b^\ast\rangle\langle b^\ast|,|a\rangle\langle a|]
		\\
		\text{and}\quad
		i\partial_\tau|b\rangle\langle b|
		=&[|a^\ast\rangle\langle a^\ast|,|b\rangle\langle b|].
	\end{aligned}
	\end{equation}
	Taking the complex conjugate of the second equation and, again, comparing the right-hand sides, we now find that
	\begin{equation}
		\hat C=|b^\ast\rangle\langle b^\ast|-|a\rangle\langle a|
	\end{equation}
	is time-independent.
	This means that $|b\rangle\langle b|=\hat C^\ast-|a^\ast\rangle\langle a^\ast|$ expresses the solution for the second subsystem, and that the first subsystem also evolves according to Eq. \eqref{eq:AppAliceOnlyEoM}, however, with a slightly modified $\hat C$.

	Finally, both cases are expressed through similar equations of motions and a time-independent operator of the form
	\begin{equation}
		\hat C=|b_0\rangle\langle b_0|+\lambda|a_0\rangle\langle a_0|,
	\end{equation}
	where the involved and normalized vectors correspond to the (complex conjugates of) states for $\tau=0$.
	With the general operator $\hat C$, and for $\lambda=1$, the separable evolution via $\hat V$ is obtained, and the settings $\lambda=-1$ and $|b_0\rangle\mapsto|b_0^\ast\rangle$ apply to the generator $|\Phi\rangle\langle\Phi|$ of the separable evolution.
	Because of the two-dimensional image of $\hat C$, we expand
	\begin{equation}
		|a(\tau)\rangle=c(\tau)|a_0\rangle+s(\tau)|b_0\rangle.
	\end{equation}
	Clearly, the initial values are $c(0)=1$ and $s(0)=0$.
	Rather than the von~Neumann equation, we can analogously solve the Schr\"odinger-type equation \cite{SW17}
	\begin{equation}
		i\partial_t|a(\tau)\rangle=\hat C|a(\tau)\rangle.
	\end{equation}
	This kind of equation of motion in a two-dimensional Hilbert space is a common study problem and yields the solutions
	\begin{equation}
	\begin{aligned}
		s(\tau)=&
			-i\frac{q^\ast}{\Delta}\sin(\tau\Delta)
		e^{-i\tau(1+\lambda)/2}
		\quad\text{and}
		\\
		c(\tau)=&\left(
			\cos(\tau\Delta)
			+i\frac{1-\lambda}{2\Delta}\sin(\tau\Delta)
		\right)
		e^{-i\tau(1+\lambda)/2},
	\end{aligned}
	\end{equation}
	where the global phase can be ignored.
	The other parameters are $q=\langle a_0|b_0\rangle$ and $\Delta=\{([1-\lambda]/2)^2+\lambda|q|^2\}^{1/2}$.

	For our specific parameters, $\lambda=\pm1$, the above general formulas simplify as follows.
	The case $\lambda=1$ for $\hat V$ yields
	\begin{equation}
		s(\tau)=-i\frac{q^\ast}{|q|}\sin(|q|\tau)
		\quad\text{and}\quad
		c(\tau)=\cos(|q|\tau),
	\end{equation}
	where the global phase was suppressed also.
	Similarly, the scenario with $\lambda=-1$ for $|\Phi\rangle\langle\Phi|$ results in
	\begin{equation}
	\begin{aligned}
		s(\tau)=&
			-iq^\ast\frac{\sin\left(\sqrt{1-|q|^2}\tau\right)}{\sqrt{1-|q|^2}}
		\quad\text{and}
		\\
		c(\tau)
		=&\cos\left(\sqrt{1-|q|^2}\tau\right)
		+i\frac{\sin\left(\sqrt{1-|q|^2}\tau\right)}{\sqrt{1-|q|^2}}.
	\end{aligned}
	\end{equation}

	The relations derived in this section then provide a complete framework to explore the spin-spin correlations (Sec. \ref{subsec:TwoQubits}) and energetically separated ground state (Sec. \ref{subsec:TwoQudits}) in the main part.

\section{Solution for the multipartite example}
\label{app:multipartite}

	We here consider a generator (i.e., rescaled Hamiltonian) that reads
	\begin{equation}
		\hat G=\Gamma\hat A^{\otimes N}+\Gamma^\ast\hat A^{\dag \otimes N},
	\end{equation}
	with $\hat A=|0\rangle\langle 1|$.
	Note that a local flip operation gives us the Hermitian conjugate of this operator, $\hat A^\dag=\hat \sigma_x\hat A\hat \sigma_x$, allowing us to assume---without a loss of generality---that terms like $\hat A^{\otimes (N-K)}\otimes\hat A^{\dag \otimes K}$ can be replaced by similar terms for $K=0$.

	Using GHZ-like states, $[|0\rangle^{\otimes N}\pm (\Gamma^\ast/|\Gamma|)|1\rangle^{\otimes N}]/\sqrt{2}$ gives us the eigenvalues $\pm|\Gamma|$.
	All other eigenvalues are zero.
	For the separability eigenvalue equation, we expand local states as $|a_j\rangle=c_j|0\rangle+s_j|1\rangle$, where $|c_j|^2+|s_j|^2=1$.
	With that, the reduced operators become
	\begin{equation}
		\hat G_{a_1,\ldots, a_{j-1},a_{j+1},\ldots,a_N}
		{=} \Gamma\prod_{k:k\neq j}c_k^\ast s_k |0\rangle\langle 1|
		{+}\Gamma^\ast\prod_{k:k\neq j}s_k^\ast c_k |1\rangle\langle 0|.
	\end{equation}
	When this operator is not zero, the eigenstates of this operator are of the form $|a_j\rangle=(|0\rangle+e^{i\varphi_j}|1\rangle)/\sqrt{2}$, i.e., $|c_j|=|s_j|=1/\sqrt{2}$.
	The separability eigenvalue equations then simplify to
	\begin{equation}
	\begin{aligned}
		&\hat G_{a_1,\ldots, a_{j-1},a_{j+1},\ldots,a_N}|a_j\rangle=g|a_j\rangle
		\\\Leftrightarrow\quad
		&\frac{\Gamma e^{i\phi}}{2^{N-1}}\frac{1}{\sqrt2}\left(
			|0\rangle+\frac{\Gamma^{\ast 2}e^{-2i\phi}}{|\Gamma|^2}e^{i\varphi_j}|1\rangle
		\right)
		\\
		=&g\frac{1}{\sqrt2}\left(
			|0\rangle+e^{i\varphi_j}|1\rangle
		\right),
	\end{aligned}
	\end{equation}
	where $\phi=\sum_{k=1}^N\varphi_k$.
	Equating coefficients, we find the constraint $\Gamma^{\ast 2}e^{-2i\phi}/|\Gamma|^2=1$, which holds true when $e^{i\phi}=\pm \Gamma^\ast/|\Gamma|$ applies.
	This results in separability eigenvalues $g=\Gamma e^{i\phi}/2^{N-1}=\pm|\Gamma|/2^{N-1}$.
	These values are then used in the main part, Sec. \ref{subsec:MultipleQubits}, to determine speed limits.

\end{document}